\documentclass[twocolumn,showpacs,preprintnumbers,amsmath,amssymb,prl]{revtex4}
\usepackage{amssymb}
\usepackage{epsfig,amsmath,graphics}
\usepackage{epstopdf}

\newcommand{\eff}{\text{eff}}
\newcommand{\rad}{\text{rad}}

\newcommand{\Rb}{\text{Rb}}

\newcommand{\SNR}{\text{SNR}}

\begin{document}

%\preprint{APS/123-QED}

\title{Testing General Relativity with Atom Interferometry}

\author{Savas Dimopoulos}
%\email{savas@stanford.edu}
\author{Peter W. Graham}
\author{Jason M. Hogan}
\author{Mark A. Kasevich}
%\email{kasevich@stanford.edu}
\affiliation{%
Department of Physics, Stanford University, Stanford, California
94305
}%

\date{\today}% It is always \today, today,

\begin{abstract}
The unprecedented precision of atom interferometry will soon lead
to laboratory tests of general relativity to levels that
will rival or exceed those reached by astrophysical observations.
We propose such an experiment that will initially test the
equivalence principle to 1 part in 10$^{15}$ (300 times better than the current limit), and 1 part in $10^{17}$ in the future. It will also probe general
relativistic effects---such as the non-linear three-graviton
coupling, the gravity of an atom's kinetic energy, and the falling of light---to several
decimals.  In contrast to astrophysical observations, laboratory tests can isolate these effects via their different functional dependence on experimental variables.
\end{abstract}

\pacs{04.80.-y, 04.80.Cc, 03.75.Dg}% PACS, the Physics and Astronomy
\maketitle

%Experimental  tests of general relativity (GR) have gone through two major phases.  The original tests of the perihelion precession and light bending were followed by a golden era from 1960 until today (for a review see \cite{Will}).
%GR tests were in part motivated by alternatives to Einstein's theory, such as Brans-Dicke \cite{Brans:1961sx}.  More recently, the cosmological constant problem suggests that our understanding of general relativity is incomplete, motivating a number of proposals for modifying gravity at large distances \cite{ccpgrmodify}. In addition, alternatives to the dark matter hypothesis have led to theories where gravity changes at slow accelerations or galactic scales \cite{dmgrmodify}.

Experimental  tests of general relativity (GR) have gone through two major phases.  The original tests of the perihelion precession and light bending were followed by a golden era from 1960 until today (see e.g. \cite{Will}).
These tests were in part motivated by alternatives to GR, such as Brans-Dicke \cite{Brans:1961sx}.  More recently, the cosmological constant problem suggests that our
understanding of gravity is incomplete, motivating a
number of proposals for modifying GR at large distances
\cite{ccpgrmodify}.  Also, alternatives to the
dark matter hypothesis have led to theories where gravity changes
at slow accelerations or galactic scales
\cite{dmgrmodify}.

Presently, most GR parameters have been tested at the part per
thousand level.
%(see Table \ref{Tab: PPN}).
%This is in contrast to QED which has been tested to many decimals.
Typical GR tests involve the study of planets, stars, or light
moving over astronomical distances for extended periods of time.
In this letter we argue that high-precision tests
of GR are possible in the lab using the motion of individual,
quantum-mechanical atoms moving over short distances for brief
periods of time. Specifically, atom interferometry can lead to laboratory tests of GR that are competitive with astrophysical tests and potentially superior to them in the
long run. There are two reasons for this: one is the high
accuracy of atomic physics methods -- reaching, for example,
16-decimal clock synchronization \cite{Oskay}.
%The second is that several control parameters can be varied in such an experiment, allowing us to isolate and measure individual relativistic terms and control backgrounds by using their different scalings with these parameters.
The second is that such an experiment has several control parameters, allowing us to isolate and measure individual relativistic terms and control backgrounds by using their scalings with these parameters.
In contrast, with astrophysical observations we have limited control and often
cannot disentangle the relativistic effects.

Our proposed experiment relies on light-pulse atom interferometers.  These have already achieved extreme accuracy as inertial sensors in a variety of
configurations including gyroscopes \cite{PhysRevLett.78.2046},
gradiometers \cite{PhysRevLett.81.971}, and gravimeters
\cite{0026-1394-38-1-4}.  The first generation of atom inteferometry experiments will push the
limits on the Principle of Equivalence (PoE) and begin measuring GR effects and placing
constraints on parametrized post-Newtonian (PPN) parameters, as shown in the third column of Table \ref{Tab: PPN}.  The next three columns show further possible improvements.

\begin{table}
\begin{center}
\begin{tabular}{|c|ccccc|}
\hline
Tested & current & AI & AI & AI & AI far\\
Effect & limit & initial & upgrade & future & future\\
\hline
PoE & $3 \times 10^{-13}$ & $10^{-15}$ & $10^{-16}$ & $10^{-17}$ & $10^{-19}$\\
PPN ($\beta$, $\gamma$) & $10^{-4}$-$10^{-5}$ & $10^{-1}$ & $10^{-2}$ & $10^{-4}$ & $10^{-6}$\\
\hline
\end{tabular}
\caption{\label{Tab: PPN} Experimental precision for measuring GR effects, PPN parameters and Principle of Equivalence (PoE) violations.  The initial atom interferometer (AI) limits assume the 10m experiment described in the text.  The potential upgrade would implement $200 \hbar
k$ LMT beamsplitters. The future experiment assumes a 100m
interferometer.  A possible improvement using Heisenberg
statistics is shown in the far future column. All precision values
assume $10^{6} \text{s}$ of integration. }
\end{center}
\end{table}

To calculate the effect of GR corrections to Newtonian gravity on
an atom interferometer, we need the metric governing the motion of
the atoms and photons in the experiment. Consider a Schwarzschild spacetime in the PPN expansion ($\hbar = c = 1$):
\begin{equation}
\label{Eqn: metric} ds^2 = (1 + 2 \phi + 2 \beta \phi^2) dt^2 -
(1-2 \gamma \phi) dr^2 - r^2 d\Omega^2
\end{equation}
where $\phi = -\frac{GM}{r}$ is the Earth's gravitational
potential, and $\beta$ and $\gamma$ are PPN parameters.  The major effect neglected here
is the rotation of the Earth.  The Newtonian effects of the
rotation are an important background, but the possible rotation related GR effects
are undetectable in the interferometer configurations considered
here.

Combining the geodesic equations for the spatial $\vec{x}^i$ ($i=1,2,3$) and $t$, the coordinate acceleration of an atom in the frame of Eqn. \ref{Eqn: metric} (with $\vec{v} = \frac{d\vec{x}}{dt}$) is
\begin{equation}
\label{Eqn: force} \frac{d \vec{v}}{dt} = - \vec{\nabla} ( \phi +
(\beta + \gamma) \phi^2 ) +
\gamma (3 (\vec{v} \cdot \hat{r})^2 - 2 \vec{v}^2) \vec{\nabla} \phi + 2 \vec{v} (\vec{v} \cdot \vec{\nabla} \phi)
\end{equation}
This illustrates two classes of
leading GR corrections to the Newtonian force law. The $\nabla \phi^2$ terms are related to the non-linear (non-Abelian) nature of gravity indicating that gravitational energy gravitates through a three-graviton vertex.
To see this, note the divergence of the gravitational field given in Eqn. \ref{Eqn: force} is nonzero because of these terms.  Just as for an electric field, a nonzero divergence of the gravitational field implies a local source density, in this case a local energy density, proportional to that divergence $\nabla \cdot \nabla \phi^2 = 2 (\nabla \phi)^2$.  So the local energy density is proportional to the field squared.  This energy gives rise to the $\nabla \phi^2$ terms.
The other terms are velocity dependent forces related to
the gravitation of the atom's kinetic energy.  The non-linear GR corrections are smaller than Newtonian gravity by a factor of $\phi \sim 7 \times 10^{-10}$ while the
velocity dependent forces are smaller by $v^2 \sim 10^{-15}$ for
the atom velocities we are considering.
%We will show that the non-linear terms can only be measured through a gradient of the force produced and so are reduced by an additional factor of $\sim 10^{-6}$, the ratio of the size of the experiment to the radius of the Earth.  Both effects are then $\sim 10^{-15} g$.
We will show that the non-linear terms can only be measured through a gradient of the force produced and so are reduced by an additional factor of $\frac{10 \text{m}}{R_\text{earth}} \approx 10^{-6}$ for a 10m long experiment.  Both effects are then $\sim 10^{-15} g$.

\begin{figure}
\begin{center}
\includegraphics[width=\columnwidth]{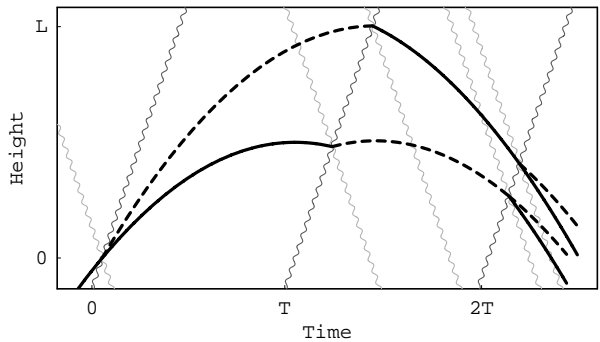}
\caption{ \label{Fig:space-time} A space-time diagram of a light
pulse atom interferometer. The black curves indicate the geodesic
motion of a single atom near the surface of the Earth.
Laser light used to manipulate the atom is incident from above
(light gray) and below (dark gray) and travels along null
geodesics.  The finite speed of the light has been
exaggerated. }
\end{center}
\end{figure}

{\it Experimental Setup.}--To measure these small accelerations we
consider first a one dimensional gravimeter arranged to measure
vertical accelerations with respect to the Earth. In a
ground-based atom interferometer gravimeter, a dilute ensemble of
cold atoms is launched upward with velocity $v_L$. The atoms then
follow trajectories in accordance with Eq. (2). During their
free-fall, a sequence of laser pulses along the vertical direction
serve as beamsplitters and mirrors by coherently transferring
photon recoil kicks of momentum $\hbar k_{\eff}$ to each atom
\cite{Berman}. For example, the laser pulse (gray) at $t=0$ in
Fig. \ref{Fig:space-time} acts as a beamsplitter, putting the atom
in a superposition of the initial velocity state (solid black) and
a state with higher velocity (dashed black). The resulting spatial
separation of the halves of the atom is proportional to the
interferometer's acceleration sensitivity. We consider the
beamsplitter-mirror-beamsplitter
$(\frac{\pi}{2}-\pi-\frac{\pi}{2})$ sequence
\cite{PhysRevLett.67.181}, the simplest implementation of a
gravimeter and the matter-wave analog of a Mach-Zender
interferometer.

To test GR, we propose a Rb interferometer with an initial precision of $\sim 10^{-15}g$.
%This will be achieved with an evaporatively cooled atom source, large momentum transfer (LMT) beamsplitters, and by launching the atoms inside a $L \approx 10\text{m}$ tall vacuum system allowing a long interrogation time $T=1.34\text{s}$.
This will be achieved with an evaporatively cooled atom source, large momentum transfer (LMT) beamsplitters, and an $L \approx 10\text{m}$ tall vacuum system for the atoms' flight, allowing a long interrogation time $T=1.34\text{s}$.
To reduce technical noise, including laser phase and vibrational noise, differential strategies similar to those used in current gravity gradiometers will be employed.  For example, to
test the PoE, the differential acceleration between $^{85}\Rb$ and $^{87}\Rb$ can be measured in a simultaneous dual species fountain \cite{Marion}.

Evaporatively cooled, rather than laser cooled, atomic sources
\cite{atomicsources} should enable a significant performance
improvement over the previous generation of sensors.  First,
evaporatively cooled sources enable tighter control over systematic
errors related to the initial position and velocity of the atomic
ensemble. Second, evaporatively cooled sources achieve temperatures
($<$ 1$\mu$K) low enough to implement LMT beamsplitters, which are
highly velocity selective. Promising LMT beamsplitter candidates
include optical lattice manipulations \cite{Phillips2002:JPhysB},
sequences of Raman pulses \cite{McGuirk} and adiabatic passage
methods \cite{Chu}. Finally, the low temperatures available with
these sources eliminate signal losses due to expansion of the
ensemble over long interferometer interrogation times.

Overall sensitivity is both a function of the effective momentum
transfer of the atom optics ($\hbar k_{\eff}$) and the
signal-to-noise ratio (SNR) of the interference fringes. Due to
technical advances in implementation of normalized detection methods
for atomic clocks and sensors \cite{clocks}, interference fringes
can now be acquired with high SNRs limited only by quantum
projection noise (atom shot-noise) for ensembles of up to $\sim
10{^7}$ atoms. Using $10\hbar k (=\hbar k_\eff)$ LMT beamsplitters,
a $\sim 10^7$ atom evaporatively cooled source and an interrogation
time of $T=1.34\text{s}$ results in a sensitivity of $7\times
10^{-13}g/ \text{Hz}^{1/2}$, and in a precision of $\sim 10^{-15}g$
after a day of integration.  This estimate is based on
realistic extrapolations from current performance levels, which are
at $10^{-10}g$ \cite{Fixler}.

{\it Interferometer Phase Shifts.}--The total phase shift in the interferometer is the sum of
three parts: the propagation phase, the laser interaction phase,
and the final wavepacket separation phase \cite{CCT:1994}. The usual formulae for these must be modified in GR to be coordinate invariants.  Our calculation will be discussed in greater detail in \cite{upcoming}.  The space-time paths of the atoms and lasers are geodesics of Eqn. \ref{Eqn: metric}.  The propagation phase is
\begin{equation}
\phi_\text{propagation} = \int L dt = \int m ds
\end{equation}
where $L$ is the Lagrangian and the integral is along the atom's geodesic.  The separation
phase is taken as
\begin{equation}
\phi_\text{separation} = \int \overline{p}_\mu dx^\mu \sim
\overline{E} \Delta t - \vec{\overline{p}} \cdot \Delta \vec{x}
\end{equation}
where, for coordinate independence, the integral is over the null
geodesic connecting the classical endpoints of the two arms of the
interferometer, and $\overline{p}$ is the average of the classical
4-momenta of the two arms after the third pulse.  The laser phase
shift due to interaction with the light is the constant
phase of the light along its null geodesic, which is its phase at
the time it leaves the laser.
Corrections due to the atomic wavefunction size $\Delta x$ \footnote{This stationary phase approximation neglects accelerations of size $(\partial_r^2 g) \Delta x^2 \lesssim 10^{-20} g$ for $\Delta x \lesssim 1 \text{mm}$.} and the laser pulse time can be calculated nonrelativistically \cite{quantum_calculation} but do not affect the leading order GR signal.
%Terms neglected by the stationary phase approximation, equivalent to neglecting the finite size of the atom's wavefunction, and assuming a negligible pulse time for the laser are relevant backgrounds and have been calculated [] but do not affect our calculation of the leading order piece of the GR signal.
%In this simplified treatment we make the stationary phase approximation, which is equivalent to ignoring the finite size of the atom wavepacket. However, effects due to finite wavepacket size and the finite pulse time of the light are included in the full analysis and do not contribute to the leading order GR signal.

%The error in this stationary phase approximation to the true quantum mechanical calculation can be estimated as the second gradient of the gravitational field (the largest effect which is not linear over the width of the atom) times the size of the atomic wavefunction squared $\sim 1 \text{mm}^2 \times (\partial_r^2 g) \sim 10^{-20} g$.
%The full calculation will be discussed in greater detail in \cite{upcoming}.

\begin{table}
\begin{center}
\begin{tabular}{|l|c|c|c|}
\hline
& Phase Shift & Size (rad) & Interpretation \\
\hline
1. & $- k_{\eff} g T^2$ & $3 \times 10^8$ & gravity \\
2. & $- k_{\eff} (\partial_r g) T^3 v_L$ & $-2 \times 10^3$ & 1st gradient \\
3. & $-3 k_{\eff} g T^2 v_L$ & $4 \times 10^1$ & Doppler shift \\
4. & $(2 - 2 \beta - \gamma) k_{\eff} g \phi T^2$ & $2 \times 10^{-1}$ & GR \\
5. & $- \frac{7}{12} k_{\eff} (\partial_r^2 g) T^4 v_L^2$ & $8 \times 10^{-3}$ & 2nd gradient \\
6. & $-5 k_{\eff} g T^2 v_L^2$ & $3 \times 10^{-6}$ & GR \\
7. & $(2 -2 \beta - \gamma) k_{\eff} \partial_r (g \phi) T^3 v_L$ & $2 \times 10^{-6}$ & GR 1st grad \\
8. & $-12 k_{\eff} g^2 T^3 v_L$ & $-6 \times 10^{-7}$ & GR \\
\hline
\end{tabular}
\caption{\label{Tab: phases} A partial phase shift list.  The
sizes of the terms assume the initial design, sensitive to accelerations $\sim 10^{-15} g$.}
\end{center}
\end{table}

In Table \ref{Tab: phases} we list some of the phase shifts that arise from an analytic
relativistic calculation.  Effectively, the local
gravitational acceleration is expressed as a Taylor series in the
height above the Earth's surface.  The first phase shift in Table
\ref{Tab: phases} represents the effect of the leading order (constant) piece of
the local acceleration while the 2nd and 5th terms are the
next gradients in the Taylor series.  Notice that even the second
gradient of the gravitational field is relevant for this
interferometer.  The 3rd term arises from the second order
Doppler shift of the laser's frequency as seen by the moving atom.
The 4th, 6th, 7th, and 8th terms arise only from GR and are not
present in a Newtonian calculation.  The 4th and
7th terms arise in part from the non-linear nature of gravity.

The 6th term receives contributions from the velocity-dependent forces in Eqn. \ref{Eqn: force}, but its coefficient is independent of $\gamma$.  There are two canceling contributions to this term coming from the $\gamma$ terms in the force law for the atom and the photon.  This term thus measures the effect of gravity on light and the velocity-dependent force on the atom.  If we put a different parameter, $\delta_\text{light}$, in front of the $\phi$ in the component $g_{00}$ of the metric governing the motion of the light, we would see this term as $(4+ \delta_\text{light} + \gamma_\text{light} - \gamma_\text{atom}) k_\eff g T^2 v_L^2$, where the $\gamma$'s are the PPN parameters in the metrics for the light and the atom.  This term then tests a matter-light principle of equivalence, namely that they both feel the same metric.

Previous works on GR and interferometry \cite{previous} have not dealt with a specific, viable experiment or a full relativistic calculation.  Important effects, such as the influence of gravity on light, the corresponding changes to the separation and laser phases, and the non-linearity of gravity were not discussed.  The typical perturbation theory calculation, which integrates the linearized GR Lagrangian over the unperturbed Newtonian trajectories, does not give the correct coefficients or even the dependence on PPN parameters of the phase shifts in Table \ref{Tab: phases}.  For example, the aforementioned cancellation of the $\gamma$'s in $k_\eff g T^2 v_L^2$ would be missed by such a calculation.

{\it Measurement Strategies.}--To test GR, the relativistic terms
must be experimentally isolated from the total phase
shift.  Many effects contribute to the total phase shift including
the Newtonian gravity of the nearby environment, magnetic fields, and the Earth's rotation.
Many of these effects will be much larger than the GR
effects.  We employ magnetic shielding, a rotation servo to null Earth rate during the interferometer interrogation time and, if necessary, strategically placed masses to 'shim' the local
gravity field.  The local field can also be characterized to
high precision using the conventional gravity response (1st term in
table \ref{Tab: phases}) at shorter times $T$.

To pick out the relevant terms, four different control
parameters can be used: $k_{\eff}, v_L$, $T$, and $R$, the distance from the center of the Earth.  Also, with a different design, the angle of the interferometer can be varied.  These
parameters can be varied by order 1, except for $R$ which can only be varied
by $\sim 10^{-6}$ in a ground based experiment.  The different scalings of the relativistic terms with these parameters allow many backgrounds to be rejected.

For example, the 6th term in Table \ref{Tab:
phases}, $k_\eff gT^2 v_L^2$, scales differently with the control parameters than the backgrounds and so can be directly measured.  Practically, to reduce systematics and technical noise, this would require a differential measurement where clouds of atoms are launched simultaneously with different velocities.

The largest GR term, $k_\eff gT^2 \phi$, is more difficult
to measure.  Its scalings with the three main control
parameters, $k_\eff, v_L,$ and $T$ allow most backgrounds to be
ignored.  However, it must still be picked out of the Newtonian background, $k_\eff gT^2$. It
originates in part from the $\nabla \phi^2$ term in Eqn. \ref{Eqn: force}
and is thus due to the non-linear nature of gravity.
Since it is impossible to have a force that scales as $\frac{1}{R^3}$ outside the mass distribution in the center of mass frame in Newtonian gravity, the $R$ scaling of this term may provide a way to distinguish it from a Newtonian gravitational field.  A differential
measurement of the $R$ scaling would involve two
interferometers running simultaneously at different heights. The
same laser should be used for both interferometers to reduce
systematics and technical noise.  At best, such a differential setup would measure
the gradient of this term, thus reducing it to $\sim 10^{-7} \rad$
or $\sim 10^{-15} g$, if the interferometers are placed 10 m
apart.

In practice, a measurement of this term or of the third GR term, $k_{\eff} \partial_r (g \phi)
T^3 v_L$, could be masked by the Newtonian
gravity of local mass inhomogeneities.  One possible approach to rejecting the
Newtonian background is to employ three such differential
accelerometer measurements along three mutually orthogonal axes.
If these three measurements are added together, all Newtonian
gravity gradient contributions to the measurement must cancel, since, in the absence of a local source, the divergence of the gravitational field is zero in Newton's theory.
In GR this divergence is not zero, as can be seen from the $\nabla
\phi^2$ term in Eqn. \ref{Eqn: force}.
%Equivalently, the gravitational field itself carries energy and so acts as a source for gravity within the experiment, thus producing a divergence of the gravitational field.
For such a strategy to be effective, however, the Newtonian field needs to be sufficiently slowly varying over the measurement baseline, which can be tested by varying the baseline.

Table \ref{Tab: PPN} summarizes the experimental precision possible for measuring GR effects and the PPN parameters $\beta$ and $\gamma$.  The initial atom interferometry limits assume the 10m experiment described earlier, but many upgrades are possible. For example, using
$200 \hbar k$ beamsplitters increases the sensitivity by a factor of $\sim 10$.  Expanding to a $100 \text{m}$ long interferometer would increase the sensitivity by a factor of 10.  For the GR terms this improvement would be even larger; for example, terms 6 and 7 scale as $L^2$ ($v_L \sim T$, $L \sim T^2$).  With these improvements it is possible to reach limits
on the PPN parameters of $\sim 10^{-4}$, competitive with present limits.  Finally, the noise
performance can be improved by using entangled states instead of
uncorrelated atom ensembles \cite{spin_squeezing}.  For a suitably
entangled source, the Heisenberg limit is $\SNR \sim n$, a
factor of $\sqrt{n}$ improvement.  For $n\sim 10^{6}$ entangled atoms, the
potential sensitivity improvement is $10^3$.  Progress using these techniques may soon make improvements in $\SNR$ on the order of 10 to 100 realistic \cite{Tuchman_PRA}.

{\it Discussion.}--By combining a long interrogation time, LMT beamsplitters
and Heisenberg statistics, a ground-based interferometer could
exceed the precision of present astrophysical tests of
GR. Even at comparable precisions, laboratory tests provide
an important complement  to astrophysical ones. An advantage of laboratory tests is that they
can isolate specific GR effects, such as the
non-linear coupling and the gravitation of kinetic energy, that
are not isolated in astrophysical tests.  For example,
the Lunar Laser Ranging (LLR) test of the PoE
constrains the PPN parameters $\beta$ and $\gamma$, but it
cannot test for the existence of the non-linear coupling. In Newton's theory the Weak Equivalence Principle (in the sense of equal Earth and Moon accelerations
towards the Sun) is also satisfied and there is no non-linear
coupling at all. Only in the subclass of deviations from GR given
by the PPN expansion does LLR imply a non-linear coupling equal
to that predicted by GR to 3 parts in $10^4$.  This happens
because PPN is a restricted class of deviations that does not
include many theories.  Using atom interferometers, the non-linear coupling can be
directly and unambiguously measured, allowing us to test a truly
relativistic effect that does not occur in Newton's theory.

Finally we consider whether the Hubble expansion rate, $H$, is measurable through the force it exerts on separated atoms. Unfortunately, according to the PoE, local experiments feel the rest of the Universe through its tidal forces proportional to Riemann $\sim H^2$, which is too small. In particular, in GR there is no local effect linear in $H$, as is sometimes invoked to explain the Pioneer anomaly \cite{pioneer}.
%Conversely, a linear dependence in $H$ would be doubly interesting as it would indicate both a violation of GR and a local measurement of $H$.
Similarly, the PoE prevents the detection of our free-fall towards the galactic dark matter in a local experiment.

%The coming generation of atom interferometers can test GR effects including the non-linearity of gravity, the gravitation of kinetic energy, the falling of light, and the PoE.  Testing the non-linearity of gravity may require a multi-axis experiment.

We thank Robert Wagoner and Nemanja Kaloper for very valuable discussions.

\end{document}